\documentclass[12pt]{article}
\def\fun#1#2{\lower3.6pt\vbox{\baselineskip0pt\lineskip.9pt
\ialign{$\mathsurround=0pt#1\hfill##\hfil$\crcr#2\crcr\sim\crcr}}}

\renewcommand\({\left(}
\renewcommand\){\right)}

\newcommand\eq[1]{Eq.~(\ref{#1})}
\newcommand\eqs[2]{Eqs.~(\ref{#1}) and (\ref{#2})}
\newcommand\eqss[3]{Eqs.~(\ref{#1}), (\ref{#2}) and (\ref{#3})}

\newcommand\ee{\end{equation}}
\newcommand\be{\begin{equation}}
\newcommand\eea{\end{eqnarray}}
\newcommand\bea{\begin{eqnarray}}

\newcommand\mpl{M_{\rm P}}

\newcommand{\lsim}{\mbox{\raisebox{-.9ex}{~$\stackrel{\mbox{$<$}}{\sim}$~}}}

\newcommand{\gsim}{\mbox{\raisebox{-.9ex}{~$\stackrel{\mbox{$>$}}{\sim}$~}}}

\def\dslash{\not{\hbox{\kern-2pt $\partial$}}}
\def\Dslash{\not{\hbox{\kern-4pt $D$}}}
\def\Oslash{\not{\hbox{\kern-4pt $O$}}}
\def\Qslash{\not{\hbox{\kern-4pt $Q$}}}
\def\pslash{\not{\hbox{\kern-2.3pt $p$}}}
\def\kslash{\not{\hbox{\kern-2.3pt $k$}}}
\def\qslash{\not{\hbox{\kern-2.3pt $q$}}}

\newtoks\slashfraction
\slashfraction={.13}
\def\slash#1{\setbox0\hbox{$ #1 $}
\setbox0\hbox to \the\slashfraction\wd0{\hss \box0}/\box0 }

\def\call{{\cal L}}

\newcommand\sub[1]{_{\rm #1}}

\newcommand\mgravvac{m_{3/2}}

\newcommand\meffs{m\sub{eff}^2}

\begin{document}

\begin{titlepage}
    
\begin{center}

\hfill TU-708\\
\hfill hep-ph/0402174\\
\hfill February, 2004\\

\vskip .4in

{\Large \bf

The masses of weakly-coupled scalar fields in the early Universe

}

\vskip .4in

{\large
David H. Lyth$^*$ and Takeo Moroi$^\dag$
}

\vskip .3in

$^*${\it Physics Department, Lancaster University, Lancaster LA1 4YB,
UK}

\vskip .2in

$^\dag${\it Department of Physics, Tohoku University, Sendai 980-8578,
Japan}

\end{center}

\vskip .5in

\begin{abstract}
    
We consider the effective mass-squared in the early Universe, of a
scalar field which has only Planck-suppressed couplings with light
fields and whose true mass is less than the Hubble parameter $H$. A
detailed investigation shows that the effective mass-squared
generically is of order $\pm H^2$ during inflation and
matter-domination, but much smaller during radiation domination. We
consider the special circumstances under which the mass-squared may be
much bigger or much smaller than the generic value.

\end{abstract}

\end{titlepage}

\renewcommand{\thepage}{\arabic{page}}
\setcounter{page}{1}

\section{Introduction}

In the early Universe, the effective mass-squared of a scalar field
$\phi$ will not necessarily be equal to its value in the vacuum (the
true value).  This is because the Lagrangian will contain an
interaction term of the form $f\phi^2$, where $f$ is a function of the
fields in the effective field theory.  In the vacuum, the mass-squared
is
\begin{eqnarray}
    m_\phi^2 = m_{\phi 0}^2 + \langle f \rangle
    \,,
\end{eqnarray}
where the first term comes from any mass term in the Lagrangian and
the second term represents the contributions from vacuum expectation
values of fields.  In the early Universe though, the expectation value
$\langle f \rangle_t$ will differ from the vacuum value, and so will
the value of the effective mass;
\begin{eqnarray}
    \meffs = m_{\phi 0}^2 + \langle f \rangle_t
    \,.
\end{eqnarray}
In particular, if the expectation value is much bigger than $m_\phi^2$
we shall have simply
\begin{eqnarray}
    \meffs \simeq \langle f \rangle_t
    \label{delm} \,.
\end{eqnarray}

The expectation value in the early Universe may arise in a variety of
ways. There may be time-independent or slowly varying scalar fields
which are far from their vacuum values, and there may also be rapidly
oscillating scalar fields.  (Here `slow' and `rapid' refer to the
timescale used to evaluate the expectation value, which is here a
time-average. To make sense, the timescale should be at least the
inverse of the effective mass.)  After re- or pre-heating there is
certainly a gas, described by particles of spin zero or higher, all of
which contribute to the (now quantum) expectation value of $f$.

If $\phi$ has renormalizable interactions {\em and} $\phi$ is close to
zero, these interactions will give the dominant contribution to
$\langle f\rangle$.  In particular, if the interactions place $\phi$
in thermal equilibrium with fields whose effective mass is much
smaller than the temperature $T$, and $m_\phi$ is also much smaller,
then typically \cite{Dolan:qd}
\begin{eqnarray}
    m\sub{eff}^2 \sim T^2 \,.
\end{eqnarray}

If instead $\phi$ is large, any field with which it has a
renormalizable interactions will usually acquire a large mass-squared,
which will suppress its contribution to $\langle f \rangle$. For
example, if there is an interaction
\begin{eqnarray}
    V \supset \lambda\psi^2\phi^2
    \label{psiphi}
\end{eqnarray}
with a scalar field $\psi$, a large value of $\phi$ will generate a
large effective mass-squared for $\psi$ (equal to $\lambda\phi^2$)
which will drive $\psi$ to zero and eliminate the contribution of this
interaction to the effective mass-squared of $\phi$.

In this paper we suppose that only non-renormalizable interactions are
significant in determining the effective mass-squared of the field
$\phi$, and that these interactions are suppressed by the Planck scale
$M_{\rm P}$.\footnote {We are not assuming that all non-renormalizable
interactions are suppressed so strongly.  For instance, in the
Horava-Witten setup with higher-dimensional quantum gravity scale
$\Lambda \sim 10^{-2} \mpl$, the non-renormalizable interactions of
the effective four-dimensional field theory are in general suppressed
by $\Lambda \ll \mpl$, leading \cite{myhigher} in general to an
effective mass-squared of order $(\mpl/\Lambda)^2 H^2$.  But if a
field lives on a brane and gets its effective mass only from a field
living on the other brane, the relevant interactions are suppressed by
$\mpl$ leading \cite{hwmass} to the usual estimate $\meffs \sim H^2$.}
In that case the standard estimate \cite{inflation,cllsw,drt} is that
in the regime $H\gg m_\phi$
\begin{eqnarray}
    m\sub{eff}^2 \sim
    \pm  H^2
    \label{generic}
    \,,
\end{eqnarray}
where $H(t)$ is the Hubble parameter and the sign may be
time-dependent.

If present, this `mass of order $H$' will be crucial for any field
which plays a significant role in the early Universe.  For instance,
it will be crucial if $\phi$ is a field breaking Peccei-Quinn (PQ)
symmetry \cite{pqflaton}, or a field responsible for Affleck-Dine
baryogenesis \cite{Affleck:1984fy} or a field which reheats the
Universe \cite{drt,kari}.  It will also be crucial if $\phi$ is a
field responsible for the primordial density perturbation, whether the
inflaton \cite{treview,book} or some `curvaton' field
\cite{Lyth:2001nq,Moroi:2001ct,Enqvist:2001zp}.

Actually, if $\phi$ is indeed the field responsible for the density
perturbation, then its mass-squared $m_*^2$ during inflation must be
at least a factor of 10 or so less than the rough estimate
\eq{generic}.  This is because of the prediction of the spectral index
$n-1\simeq \frac{2}{3} V''_*/H_*^2$ (where $V''$ is the second
derivative of the potential at horizon exit) together with the
1-$\sigma$ estimate \cite{wmapspergel,wmapsdss} $n = 0.97 \pm 0.03$.
Furthermore, if $\phi$ is not the inflaton then its mass-squared may
need to be suppressed even after inflation \cite{dllr2} (until $H$
falls to the true mass) so as to preserve the fractional perturbation
in $\phi$.

Our main purpose here is to consider, in more detail than before, the
circumstances under which the mass of order $H$ will in fact be
generated by Planck-suppressed interactions, working as usual in the
context of supergravity.  We first consider the era of inflation,
confirming that the effective mass will generically be of order $H$.
We consider carefully the ways in which the mass of order $H$ might be
avoided for one or more fields.  Then we go on to consider the
situation after inflation, when the Universe is gaseous. In the case
of matter domination, we confirm the effective mass of order $H$,
including the case that the matter has spin greater than zero. Then we
go on to consider the case of radiation domination, which has not
previously been considered, and find that the generic mass is much
less than $H$.  (In all of these cases, the `generic' value of the
mass is coming from derivatives of the Kahler potential which are
indeed generically present. We do point out that less generic terms in
the superpotential may generate a bigger value mass.)  We conclude by
assessing the implication of our results.

\section{Preliminaries}

\subsection{Defining the effective mass-squared of a real scalar field}

If an effective field theory contains $N$ real fields $\phi_i$, the
Lagrangian of the fields is
\begin{eqnarray}
    \call = \sum G_{ij} \partial_\mu \phi_i \partial^\mu \phi_j
    -V(\phi_1,\phi_2,\cdots)
    \,.
\end{eqnarray}
The fields are coordinates in an $N$-dimensional field space with
metric $G_{ij}$. As a result, one can in a sufficiently small region
of field space, choose the fields so that $G_{ij}=\delta_{ij}$. (In
our case `sufficiently small' means `much smaller than $\mpl$', since
we are assuming that relevant non-renormalizable terms in the
Lagrangian are Planck-suppressed.)  The fields then become Cartesian
coordinates in a flat space, and are defined when both the origin and
the orientation of the axes is defined.

Usually the origin is chosen to be the vacuum expectation value (VEV)
of the fields, corresponding to a minimum of $V$. Then the
mass-squared matrix of the fields is defined as
$m_{ij}^2=\partial^2V/\partial\phi_i\partial\phi_j$, and it may be
convenient to choose the fields so that $m_{ij}^2=\delta_{ij}m^2_i$
which then defines the masses-squared $m_i^2$.

If all or some of the fields in the effective theory are charged under
a symmetry group with a fixed point, then it may be appropriate to
choose that point as the origin.  The masses-squared are defined in
the same way as before, but they can now be either positive or
negative (tachyonic). If any field has a tachyonic mass-squared, its
VEV is not at the fixed point and some symmetry is spontaneously
broken. The masses-squared defined with respect to the fixed point are
then different from those defined with respect to the vacuum.  In the
Standard Model and the Minimal Supersymmetric Standard Model (MSSM),
both definitions are used and one has to tell from the context which
one is meant. For instance, the origin is taken to be the fixed point
when one considers the running of masses-squared, from a high scale
where they are perhaps all positive to a low scale where the Higgs
masses-squared become tachyonic. If only a subset of the fields is
charged under a symmetry group with a fixed point, all of this applies
to that subset.

In some situations, one defines a field more generally as simply the
distance along a chosen line in field space.  To the extent that
motion in the perpendicular directions can be ignored, a field defined
in this way still has the dynamics of a canonically-normalized field.
A familiar case is where a complex field $\Phi \equiv (\phi_1 +
i\phi_2)/\sqrt 2$ is written in the form $\Phi=r\exp(i\theta)$, and
$r$ has a nonzero VEV $v$. Then one considers the angular field
$\phi=\sqrt 2 v \theta$.

This description of the possible ways of defining a field remains
valid in the early Universe, if `VEV' is replaced by `a
time-independent {\em or slowly-varying} value'. (The latter
possibility arises because of Hubble-damping.)  Among the additional
possibilities, we mention the definition of the inflaton field as the
distance along the slow-roll trajectory.

A useful definition of effective mass-squared for a field $\phi$, at a
given epoch in the early Universe, is simply
\begin{eqnarray}
    \meffs = V''
    \label{vpp}
    \,,
\end{eqnarray}
where a prime denotes a derivative with respect to $\phi$, evaluated
using the field values corresponding to the given epoch.  (If a field
is rapidly oscillating at the epoch, one should use its mean value.)
It often happens that the potential is usefully expanded about a
maximum,
\begin{eqnarray}
    V = V_0 - \frac12 \tilde \meffs \phi^2 + \cdots
    \label{meffsdef}
    \,,
\end{eqnarray}
with the remaining terms small until a minimum at $\phi\sub{min}$, and
$\phi$ at the epoch in question being between the maximum and the
minimum.  Then the effective mass-squared $\meffs\equiv V''$ will
typically be of order $\tilde \meffs$ in magnitude.

The origin in \eq{meffsdef} need not be a fixed point of the
symmetries. Indeed it cannot be if $\phi$ is a pseudo Nambu-Goldstone
boson (PNGB), here defined as one for which some symmetry takes the
form of a shift symmetry $\phi\to \phi+{\rm const}$.  Even if the
origin is a fixed point, $\phi$ may not be one of the fields which
diagonalise the mass matrix at the origin; that situation arises for
the flat directions of the MSSM.

\subsection{Matter fields and moduli}

A given epoch in the history of the Universe corresponds to some point
in field space, and at that epoch one can divide the fields into two
types according to whether or not they are charged under symmetries
having a fixed point which is nearby in Planck units. Adopting that
point as the origin, canonically-normalized fields of the first type
have values $\ll\mpl$ at the given epoch, while fields of the second
type have values $\gsim \mpl$.  Borrowing loosely from string
terminology we may call them respectively matter fields and moduli.
(The borrowing is loose, because the moduli space may have points of
enhanced symmetry as will be discussed in a moment.)

For the matter fields, it is useful to choose the fixed point as the
origin. Then the values of the canonically-normalized matter fields at
the given epoch are, by definition, all much less than $\mpl$. As a
result, it is enough to keep only low-order terms in the expansion of
the potential in powers of the fields.  The non-renormalizable terms
of this expansion (dimension bigger than $4$) will be suppressed by
some mass scale, and we are interested in terms suppressed by $\mpl$.

The moduli, by definition, do not transform under symmetries with a
nearby fixed point. Examples suggested by the string theory,
corresponding to canonically-normalized real fields, are as
follows. First, the dilaton which may not transform under any symmetry
at all. Next, fields corresponding to the size and shape of extra
dimensions, or the distance between branes, which during the given
epoch are at a distance of order $\mpl$ from a fixed point of their
symmetry group.\footnote
{The symmetry group of the bulk moduli is
supposed to be at least the group of modular transformations (a
discrete gauge symmetry), with possibly additional symmetries
corresponding to gauge and Yukawa interactions with either the
Standard Model sector or a hidden sector. If such additional
symmetries are present, the fixed point is said to be a point of
enhanced symmetry. If a bulk modulus is at, or close to, a fixed point
it is classified as a matter fields for the present purpose.  The
classification might change with time, so that for example inflation
takes place near a fixed point which is at a Planckian distance from
the vacuum.}
Finally, string axion fields which possess a $Z_2$ shift symmetry,
$\phi\rightarrow \phi+\Delta$ with $\Delta\sim \mpl$; in other words,
the potential of these fields is periodic with period of order $\mpl$.
(A given string axion field may be paired with a field of the first
two types, to form a complex field in a chiral supermultiplet.)  As
far as we are aware, this list exhausts the possibilities suggested by
string theory at the moment.  Other possibilities might of course come
up in the future, including ones which generate the widely-consider
`chaotic inflation' paradigm corresponding to an inflaton potential
$m^2\phi^2$ out to $\phi \sim 10\mpl$.\footnote {Such a potential has
been derived \cite{gauge} from Wilson line gauge symmetry breaking in
a large extra dimension, but from the four-dimensional viewpoint this
potential is generated by the one-loop (Coleman-Weinberg) potentials
due to an infinite number of fields with arbitrarily high mass (a
Kaluza-Klein tower). Hence it does not correspond to an effective
four-dimensional field theory of the usual kind, which by definition
contains only a finite number of fields with masses below the Planck
scale. Also, it has been suggested \cite{banks03} that this scenario
has no known string-theoretic realization.}

The contributions of the moduli to the potential depend on the version
of string theory adopted and even within a given version firm
predictions are hard to come by \cite{joebook}. However, a rather
common expectation \cite{fofphi} is that the effective potential in
the direction of a modulus $\phi$ is of the form
\begin{eqnarray}
    V(\phi) = \Lambda^4  \times f(\phi/\mpl)
    \,,
    \label{modpot}
\end{eqnarray}
with $f$ and its low derivatives of order 1 in magnitude at a generic
point within the expected range $0\lsim\phi\lsim\mpl$ of the modulus.

Another common expectation, at least in the context of the heterotic
string, is that the potential vanishes in the limit of unbroken
SUSY. In the vacuum $\Lambda$ may then be expected to be of order the
SUSY-breaking scale $M\sub S$, making the modulus mass-squared of
order the gravitino mass-squared,
\begin{eqnarray}
    |\meffs|^2\sim m_\frac32^2 =  3M\sub S^4/\mpl^2
    \,.
\end{eqnarray}
In the early Universe in the regime $\rho\gsim M\sub S^4$, the SUSY
breaking scale becomes $\rho$, which naively suggests $\Lambda\sim
\rho^\frac14$ corresponding to $|\meffs| \sim H^2$.  However, this
conclusion is not inevitable, and in particular we shall show that
during a radiation-dominated era $\meffs$ can be much smaller than
$H^2$.  Going the other way, we note also that in other string theory
schemes the moduli are supposed to be very heavy, so that even during
inflation $|\meffs|\gg H^2$.

\subsection{The supergravity potential}

The Lagrangian of $N=1$ supergravity is determined by three functions
(plus one or more Fayet-Iliopoulos constants).  These are the Kahler
potential $K$, superpotential $W$, and the gauge kinetic function. All
three are functions of the complex scalar fields $\Phi_i$, each of
which corresponds to the scalar part of a chiral supermultiplet.

Possible set of the Planck-suppressed operators depends on the model
since some the operators may be forbidden or suppressed by some
symmetry \cite{thooft}.  The superpotential and the gauge kinetic
function are holomorphic functions of the chiral superfields $\Phi_i$,
and hence it is relatively easy to find a symmetry which forbids some
class of operators.  (Of course, it is unclear if such a symmetry is
respected by the physics at the Planck scale; if not,
Planck-suppressed operators show up.)  On the contrary, it is more
difficult to control the Kahler potential since it depends on the
$\Phi_i$ and their complex conjugates $\bar\Phi_i$.

The supergravity potential is of the form
\begin{eqnarray}
    V=V_F + V_D
    \,.
    \label{vfd}
\end{eqnarray}
We shall be mostly interested in the $F$-term, which is given by
\begin{eqnarray}
    V_F =  e^{K/\mpl^2}
    \( D_i W K^{i\bar j}  \overline{D_j W} -
    3 |W|^2/\mpl^2 \)
    \label{vsugra}
    \,.
\end{eqnarray}
Subscripts for Kahler and superpotentials denote partial derivatives
with respect to corresponding fields, $K^{i\bar j}$ is the inverse of
the Kahler metric $K_{i\bar j}$, and
\begin{eqnarray}
    D_i W  \equiv W_i + W K_i/\mpl^2
    \,.
\end{eqnarray}
It is useful to separate \eq{vsugra} into two parts,
\begin{eqnarray}
    V_F =  \tilde M\sub S^2 -
    3 e^{K/\mpl^2}  |W|^2/\mpl^2
    \label{vsugra2}
    \,.
\end{eqnarray}
The first term spontaneously breaks supersymmetry, and we called it
$\tilde M\sub S^4$ because its vacuum value is usually denoted by
$M\sub S^4$. In the vacuum the two terms accurately cancel for some
mysterious reason (the cosmological constant problem).

\section{During inflation}

\subsection{The effective mass-squared during inflation}

We focus on the usual case that the inflaton potential is dominated by
the $F$-term, except for a few remarks on $D$-term inflation at the
end of the section.  In that case, it has long been known
\cite{inflation} that Planck-suppressed interactions generically give
$\meffs \sim \pm H_*^2$, for each scalar field whose true mass is less
than the inflationary Hubble parameter $H_*$.\footnote
{Such fields are
expected to exist, since according to existing inflation models $H_*$
is at least of order the gravitino mass $\mgravvac$
\cite{p03stringph}, while one may expect that at least some scalar
fields get their mass from gravity-mediated SUSY breaking
corresponding to $m_\phi\sim \mgravvac$.}
We here consider the situation more carefully than before, to better
understand the circumstances under which $|m\sub{eff}^2|$ might
instead be much bigger or much smaller than $H_*^2$.

A simple argument due to Stewart \cite{ewansugra} shows that $\meffs$
defined by \eq{vpp} is generically {\em at least} of order
$H^2$. Focusing on a particular time, we can choose the origin in
field space so that all fields vanish at that time.  Demanding that
each field is canonically normalized (part of the definition of
$\meffs$), the power series expansion of the Kahler potential in
$\Phi_i$ and $\bar\Phi_i$ must then contain the terms
\begin{eqnarray}
    K \supset \sum_i \pm |\Phi_i|^2
    \label{kterm}
    \,,
\end{eqnarray}
leading indeed to canonical normalization at the origin,
\begin{eqnarray}
    \sum K_{i\bar j} \partial^\mu \Phi_i \partial_\mu \bar \Phi_j
    = \sum  \partial^\mu \Phi_i \partial_\mu \bar \Phi_i
    \label{canonkin}
    \,.
\end{eqnarray}
Through the factor $\exp(K/\mpl^2)$ in \eq{vsugra}, the term
\eq{kterm} gives a contribution $\pm V/\mpl^2= \pm 3H^2$ to the
effective masses-squared of each field. In the generic case this
contribution will not be canceled by other contributions, leading as
advertised to $|m\sub{eff}^2|$ at least of order $H^2$.

We would like to point out that the sign in \eq{kterm} need not be
positive as is usually assumed. The sign is a property of the
effective supergravity theory, to be determined presumably by string
theory which seems to have no prejudice towards the positive sign. In
particular, a commonly occurring approximation (for instance in the
untwisted sector for the weakly coupled heterotic string) is $K\simeq
\mpl^2 \ln(1- \sum_i|\Phi_i|^2/\mpl^2)$, which gives the negative
sign.

The above argument is important because it does not invoke an
interaction term in either $K$ or $W$, relying instead on the form of
the supergravity potential to generate a particular Planck-suppressed
term in the potential $V$. Additional Planck-suppressed terms in $V$
will in general be generated by Planck-suppressed terms in $K$ and/or
$W$, some of which may contribute to the effective
mass-squared. Keeping the convention that the origin of field space
corresponds to the values of the fields at the time under
consideration, there may be terms contributing to $K$ the amount
\begin{eqnarray}
    \Delta K = \lambda_i \Phi_i^2 + {\rm c.c.}
    \,.
    \label{dhlk}
\end{eqnarray}
Taking for definiteness $\lambda_i$ either real or imaginary, these
terms contribute to the mass-squared of respectively the real and
imaginary parts of $\Phi_i$ and amount $\pm|\lambda_i| H^2$. To
understand the significance of such terms, it is helpful to transfer
them to the superpotential using the Kahler transformation which
leaves invariant the form of the supergravity potential;
\begin{eqnarray}
    K &\to&  K - \Delta K \,, \\
    W &\to& e^{\lambda_i \Phi_i^2/\mpl^2} W
    \,.
\end{eqnarray}
We see that the terms we are considering are equivalent to
Planck-suppressed interaction terms in the superpotential, which means
that we expect $|\lambda_i|\lsim 1$ as far as the cutoff scale of the
higher dimensional operators is the Planck scale. Therefore, these
terms are expected to contribute at most of order $\pm H^2$ to the
effective masses-squared.  Notice that the interaction given in Eq.\ 
(\ref{dhlk}) can be forbidden by some symmetry if $\Phi_i$ has some
non-trivial transformation property.  If the $\Phi_i^2$ term in the
Kahler potential is consistent with all the symmetry, on the contrary,
mass term of $\Phi_i$ may be also allowed in the superpotential
(unless $R$-symmetry is imposed).  Importantly, natural scale of the
mass of $\Phi_i$ in this case is of order the cutoff scale (i.e.,
$M_{\rm P}$ in our case) if there is no suppression for the
interaction of $\Phi_i$.  If so, the (effective) mass of $\phi$ is
always dominated by the superpotential contribution and the
contribution of the Kahler interaction is sub-dominant.

Now we discuss more generally the effect of Plank-suppressed
interaction terms, considering only the effective masses-squared of
the matter fields so as to take on board the possible effect of
symmetries. Our analysis goes in some respects a little beyond the one
of \cite{drt}.

To take into account the possible effect of the moduli on the
masses-squared of the matter fields, we shall assume a potential of
the form \eq{modpot} for the moduli.  Since the matter field values
are small on the Planck scale, one need keep only low-order terms in
the power series expansions of $K$ and $W$. Each term in $W$ is
holomorphic, and it is useful to adopt the convention \cite{wein3}
that $K$ has no holomorphic terms (possible since such terms can
always be transferred to $W$ by a Kahler transformation). The terms
considered in \eq{kterm} will then be the leading ones,
\begin{eqnarray}
    K =  \sum_i \pm |\Phi_i|^2  + \cdots
    \label{kterm2}
    \,.
\end{eqnarray}
Through the term $\exp(K/\mpl^2)$, these terms will generate for each
field a contribution $\pm 3H^2$ to $\tilde \meffs$, as we already
discussed.

To this contribution effect of the non-renormalizable terms in $K$ and
$W$ has to be added. To discuss them, we suppose for simplicity that a
single term $i=I$ dominates the supergravity potential \eq{vsugra},
and we also make the reasonable assumption $|K|\lsim \mpl^2$.  The
generic estimate $\tilde \meffs \sim \pm H^2$ comes from the
contribution of non-renormalizable terms in $K$.  It is analogous to
the contribution which, in the vacuum, generates soft masses-squared
of order $\pm M\sub S^4/\mpl^2$.  During inflation, the corresponding
contribution is $\pm \tilde M\sub S^4/\mpl^2$, which indeed is
expected to be of order $\pm H^2$ because the terms of \eq{vsugra2}
are not expected to cancel as they do in the vacuum.

The field $\phi$ whose mass-squared we are estimating will generally
be either the radial or the angular component of some complex field
$\Phi$.  Except for some remarks at the end, we take it to be the
radial component.  Assume first that $\Phi$ is charged only under
$U(1)$ symmetries, which act on its phase. None of these symmetries
can prevent a contribution
\begin{eqnarray}
    K^{I \bar I} \supset \lambda |\Phi|^2 / \mpl^2
    \label{kibari}
    \,.
\end{eqnarray}
The coupling $|\lambda|$ is generically expected to be of order 1,
giving indeed a contribution of order $\pm H^2$ to $\tilde \meffs$.
If $\Phi_I$ is a matter field, the contribution \eq{kibari} comes from
a quartic term in the power series expansion,
\cite{Gaillard:1993es,Bagger:1995ay,drt}
\begin{eqnarray}
    K \supset \lambda |\Phi_I|^2 |\Phi|^2/\mpl^2
    \,.
\end{eqnarray}
If instead $\Phi_I$ is a modulus, the same estimate follows if we
invoke \eq{modpot} so that
\begin{eqnarray}
    K \supset f(\phi_I/\mpl) |\Phi|^2
    \,,
\end{eqnarray}
where $\phi_I$ is the real or imaginary part of $\Phi_I$.

If instead $\Phi$ is charged under continuous non-Abelian symmetries,
one can replace $|\Phi|^2$ by an invariant sum $\sum_i |\Phi_i|^2$ or
one can replace $\Phi$ by an invariant product of fields.  Then the
term \eq{kibari} is, again, not forbidden by any symmetry. On the
other hand, if the non-abelian symmetry is global it still makes sense
to consider $\Phi$ itself, which becomes a PNGB if the global symmetry
is spontaneously broken \cite{Stewart:2000pa}.

If we make the reasonable assumption that the second term of
\eq{vsugra2} is actually negligible, with also $|K_I|\lsim \mpl$, then
the contribution \eq{kibari} to $\tilde \meffs$ is actually the only
one coming from Planck-suppressed terms in $K$. Aside from this
contribution, the effective masses-squared may receive contributions
of order $\pm H^2$ from the terms we discussed after \eq{dhlk}, but
since this can be regarded as a contribution to $W$ it may be
forbidden by symmetries for the matter fields that we are now
considering.

Typically, there are no other contributions to the effective
masses-squared which are of order $\pm H^2$. An exception arises if
the superpotential is proportional to the field $I$.  Then, $\tilde
\meffs$ for the field $I$ receives additional contributions of order
$\pm H^2$.  If (i) the plus sign of \eq{kterm2} holds, and (ii)
$|K_i|\ll \mpl$ for every field \cite{treview}, these additional
contributions cancel \cite{cllsw} those from the factor
$\exp(K/\mpl^2)$, leaving only the contribution from \eq{kibari}. (The
first condition has not been noted before.) Then one can achieve
$|\tilde \meffs| \ll H^2$ by supposing that $|\lambda|$ in \eq{kibari}
is suppressed.  This idea has been widely adopted as a paradigm for
inflation \cite{treview}, but it is incompatible with the existence of
moduli having a Kahler potential of the form \eq{modpot}, since that
will give $|K_i|\sim \mpl$ leading to additional contributions
\cite{treview} to $\tilde \meffs$.

Aside from the above exception, interaction terms in $W$ give a
contribution to $\tilde \meffs$ that has nothing to do with $H^2$.
This contribution involves only fields with non-zero values, and so
may be regarded as a generalization of the Higgs effect (which, in the
vacuum, generates the true mass of a particle through its gauge
coupling to a nonzero field).  We discussed already (\eq{psiphi}) the
effect of a renormalizable term, and similar considerations apply to
non-renormalizable terms.  Consider
\begin{eqnarray}
    W \supset  \frac{\lambda}{2\mpl} \Psi^2 \Phi^2
    \label{wbit}
    \,,
\end{eqnarray}
with $|\lambda|\sim 1$, and $\Psi$ any field with a nonzero value. It
contributes to the potential terms
\begin{eqnarray}
    V \supset \lambda  |\Psi|^4| \Phi|^2 /\mpl^2 +
    \lambda  |\Phi|^4 |\Psi|^2 /\mpl^2
    \label{vbit}
    \,.
\end{eqnarray}
This will give a contribution $\meffs =\lambda |\Psi|^4/\mpl^2$, which
could be bigger than $H^2$.  However, similarly to the situation we
discussed after \eq{psiphi}, this contribution will not appear if
$|\Phi|$ is large because then $\Psi$ will have a large positive
mass-squared which drives it to zero.  In addition, notice that it may
also be possible to forbid or suppress the term \eq{wbit} using
symmetry arguments.

All of this concerns the case that $\phi$ is the radial component of
$\Phi$. We end this section by briefly considering the opposite
situation, that $\phi$ corresponds to the angular component of $\Phi$
with the radial component fixed or slowly-varying \cite{drt,cdl}.  The
term $\exp(K/\mpl^2)$ does not contribute to the $\meffs$ of the
angular field, but non-renormalizable terms in $K$ generate $\meffs$
through a generalized $A$-term. In the absence of symmetries one again
finds $\meffs\sim \pm H^2$, but now the mass can easily be suppressed
by invoking symmetries.

\subsection{Suppressing the mass-squared during inflation}

As we already noted, at least the field responsible for the primordial
curvature perturbation must have $|m^2|\lsim 0.1 H^2$, somewhat
smaller than the generic expectation. This could of course be an
accident, but there has been a lot of discussion about how instead the
suppression might be very strong, and come about for a definite
reason.  Depending on the mechanism, the strong suppression might hold
only for very special fields, for a wide class of fields or for all
fields.  We here briefly recall that discussion, emphasizing some
points which were perhaps previously obscure and some possible new
directions for research.

Considering some particular field $\phi$ with effective mass
$m\sub{eff}$, suppose first that it is a pseudo-Nambu-Goldstone-boson
(PNGB).  This means that the Lagrangian is approximately invariant
under a global symmetry which includes the shift symmetry
$\phi\to\phi+{\rm constant}$. The mass $m\sub{eff}$ vanishes in the
limit of unbroken shift symmetry, and with sufficiently weak symmetry
breaking it will in general be less than $H$ during
inflation.\footnote
{An important exception can arise in the case of
the inflaton in a non-hybrid model (`Natural Inflation'
\cite{naturalinf}). Then the inflationary potential vanishes in the
limit of unbroken symmetry, and so does $H$.  As a result the symmetry
does not generally help to keep $m/H$ small in this case
\cite{Cohn:2000hc}. Even then though, one can have $H/m\to 0$ in an
appropriate limit of unbroken symmetry if symmetry breaking is
controlled by two independent parameters. This is what happens in the
case of gauge inflation \cite{gauge}.}

If $\phi$ is the only field affected by the shift symmetry, it can be
realized by choosing $K$ to be a function only of $\Phi+\bar\Phi$ and
$W$ to be either independent of $\Phi$ or else of the form $\exp ({\rm
const}\times\Phi)$. The axionic components of string moduli are
supposed to be examples of this. Little seems to be known about the
manner in which the shift symmetry is broken in this case
\cite{strax}, but one may hope that it will be sufficiently mild to
avoid the mass of order $H$ for the string axions.  It should be noted
that the flatness of the potential of a PNGB may be protected by the
symmetry over the entire range of field values, not just near the
origin at which the mass-squared is defined.  In other words, the
coefficients of higher powers of the field may also be
controlled. Whether this happens or not depends of course on the way
in which the symmetry is broken.

Instead of imposing a symmetry on the whole Lagrangian, one can
suppress the mass of one or more scalar fields by choosing special
forms for $K$ and $W$ which do not correspond to a symmetry, but which
{\em during inflation} keep the potential perfectly flat in the limit
that these forms hold exactly. In other words, one can demand that the
fields under consideration are effectively PNGB's during inflation.  A
general recipe for doing this was given by Stewart \cite{ewansugra},
the first step of which is to invoke an $R$-parity which keeps $W=0$
during inflation so that \eq{vsugra} takes the simple form
\begin{eqnarray}
    V = e^{K/\mpl^2} K^{i\bar j} W_i W_j^*
    \,.
\end{eqnarray}
A specific realization of this recipe in the context of the weakly
coupled heterotic string was given in \cite{glm}. This realization
works with the untwisted sector fields $\Phi_{Ii}$ and the modulus
$T_I$ (with index $I$ being specifying the modulus), and assumes that
in a sufficiently good approximation the Kahler potential depends only
on the quantities
\begin{eqnarray}
    x_I \equiv T_I + \bar T_I - \sum_i |\Phi_{Ii}|^2/\mpl^2
    \,,
\end{eqnarray}
which determine the radii of compactification of the three tori.  By
virtue of this assumption, the Kahler potential possesses the
`Heisenberg invariance' $\Phi_{Ii}\to\Phi_{Ii}+\epsilon_{Ii}$, $T_I\to
T_I + \sum_i \epsilon_{Ii}^*\Phi_{Ii}$, which would be shift
symmetries for the untwisted fields if the superpotential did not
depend on them.

The PNGB and `effective PNGB' possibilities have been explored both
for the inflaton
\cite{Cohn:2000hc,Stewart:2000pa,gauge,kklmmt,henry,kaplan,renata,%
string-inspired, shift-chaotic} (see also \cite{ars} for the indirect
use of a PNGB) and for the the curvaton
\cite{Lyth:2001nq,Dimopoulos:2003az,cdl}, and they constitute at
present a very active area of research.

Prior to \cite{glm} a different and simpler scheme was proposed
\cite{adhoc,gmo}, working with an overall modulus $T$ and $K$ a
function only of $x\equiv ( T + \bar T - \sum |\Phi_i|^2 /M_{\rm
P}^2)$ the sum running over the untwisted fields.  Unfortunately there
does not seem to be any well-motivated form for $K$ which allows this
scheme to work.  The original idea was to use the `no-scale' paradigm,
according to which $K=-3M_{\rm P}^2\ln x$, and $W$ depends only on the
untwisted fields.  This leads to
\begin{eqnarray}
    V = \frac{1}{3x^2}
    \sum | W_i |^2
    \,.
\end{eqnarray}
In the vacuum, $V=0$, and no soft masses are generated because $V$ is
the same as for unbroken global supersymmetry except for the first
factor.  (The no-scale case is not generally thought to be a good
approximation to reality, but we do not enter into that issue here.)
But during inflation, where $V$ is non-vanishing, the no-scale form
cannot work because $x$ will run away driving $V$ to zero. The
situation could be rescued if $K$ had an additional term depending
only on $x$ but existing proposals are ad hoc \cite{adhoc}, except for
that of \cite{gmo} which invokes a quite special loop correction. The
latter requires the last term of \eq{vsugra} to accurately cancel
$V_F$, and no inflation model has been exhibited in which it works.

Finally, the mass of a field may be suppressed by the quantum
correction, provided that the field has unsuppressed interactions, and
global supersymmetry is a good approximation with $m\sub{eff}$ a soft
mass. In that case, $m^2\sub{eff}$ becomes $\phi$-dependent (a running
mass) and may pass through zero.  In the vicinity of the zero, the
potential will have a maximum or minimum with $m\sub{eff}\ll H$.  This
possibility has been explored so far only for the inflaton
\cite{running1,running2,clm,ks} but it may also make sense to consider
it for the curvaton in the case that the potential develops a minimum.
(After inflation the interactions of the curvaton should be suppressed
so that it does not decay promptly, but that need not be the case
during inflation.)

We end by mentioning the case of $D$-term inflation
\cite{ewansugra,Halyo:1996pp,Binetruy:1996xj}.  If the $D$-term
dominates completely and the gauge kinetic function has negligible
dependence on the inflaton field, the inflaton potential is given by
the one-loop potential of the waterfall field (i.e., the inflaton) $S$
and there is no mass term.  One potential problem of this scenario is
that, if the superpotential or the gauge kinetic function depends on
$S$ through Planck-suppressed terms, Planck-suppressed interactions
significantly affect the inflaton potential since the inflaton field
value is of order $\mpl$ when cosmological scales leave the horizon
\cite{myhigher,km}.  One possibility of eliminating such
Planck-suppressed interactions is to introduce some symmetry under
which $S$ transforms.  Such (global) symmetry may be, however,
violated once the effects of the quantum gravity is taken into
account.  Also, with a negligible $F$ term it is difficult to see how
the dilaton can be stabilized \cite{running2}.  So it is unclear if a
realistic model of the $D$-term inflation can be really constructed.

Let us summarize this section. During inflation, the mass of order $H$
for a given field can be avoided if the field is a PNGB, or if its
mass runs, or if the potential during inflation has a special
string-inspired form.  It is also avoided if inflation occurs with the
$F$ term completely negligible ($D$-term inflation) though, in such a
case, the model should be arranged such that all the dangerous
Planck-suppressed interactions somehow vanish.  In addition, we have
also seen that, contrary to what is often stated, no-scale
supergravity does not avoid the mass of order $H$.

\section{After inflation}

Now we consider what happens after inflation.  According to present
ideas, the Universe after inflation is supposed to be gaseous save for
exceptional epochs. Possibilities for the latter include brief phase
transitions (electroweak for instance) and of course brief further
inflation (such as thermal inflation). These, though, typically occur
after $H$ falls below $m$ in which case they are irrelevant in the
present context.

More relevant is the epoch immediately after inflation, when the
field(s) responsible for the inflationary potential oscillate. In a
non-hybrid model there is just the inflaton field.  In that case, if
the field does not decay rapidly, the oscillation after at most a few
Hubble times will usually become almost sinusoidal. Then it is
equivalent to a matter-dominated gas on timescales much bigger than
the inverse effective mass of the inflaton.  Instead though, the
inflaton oscillation may efficiently create particles through
non-perturbative effects generally known as preheating.  Also, for
hybrid inflation the waterfall field oscillates and interacts with at
least the inflaton field. The situation immediately after inflation
may therefore be very complicated. However, one still expects to find
again a gas after a Hubble time or so, consisting of particles created
by the oscillation plus maybe a single oscillating field. The one
exception might be the case where the interacting inflaton and the
waterfall fields oscillate with negligible decay; we ignore this case
for the moment.

We thus proceed on the assumption that the Universe after inflation
consists of a gas and/or one or more non-interacting and oscillating
homogeneous scalar fields, the latter being equivalent to a
matter-dominated gas. This means in practice that the total energy
density is either radiation- or matter-dominated, since transitions
from one case to the other will always be rather brief.

The particle species making up the gas may have unsuppressed
interaction terms, but the particles in the gas are supposed to be
moving freely which means that the expectation values of such
interaction terms are negligible.  This means that among the terms in
the Lagrangian which involve the gas particles, only the kinetic terms
$\call\sub{kin}$ and the mass terms $\call\sub{mass}$ can have
significant expectation values.

We are interested in the effective mass-squared of a scalar field
$\phi$, which does not correspond to any of the gas particles, and
which has only Planck-suppressed interactions with those particles.
We are considering an epoch when $H$ is bigger than the true mass
$m_\phi$, and we ask whether Planck-suppressed interactions induced by
$K$ will lead to an effective mass-squared of order $\pm H^2$.

We begin by recalling some results from the theory of free fields,
which follow from the fact that the Lagrangian is essentially that of
uncoupled harmonic oscillators. First, the kinetic and mass terms of
the Lagrangian are equal by virtue of the field equations;
\begin{eqnarray}
    \langle\call\sub{kin}\rangle = - \langle\call\sub{mass}\rangle
    \label{kineqmass}
    \,.
\end{eqnarray}
Next, the energy density of the gas is of the form
\begin{eqnarray}
    \rho = \rho\sub{kin} - \langle \call\sub{mass} \rangle
    \,,
\end{eqnarray}
the first term involving spacetime derivatives.  For radiation
domination the mass term is negligible so that
\begin{eqnarray}
    | \langle \call\sub{mass} \rangle | \ll  \rho = 3H^2 \mpl^2
    \label{raddom}
    \,.
\end{eqnarray}
For matter domination, spatial derivatives give a negligible
contribution to $\rho\sub{kin}$, and as a result $\rho\sub{kin} =
\langle \call\sub{kin}\rangle$, leading to
\begin{eqnarray}
    | \langle \call\sub{mass} \rangle | = \frac12\rho
    =\frac32 H^2\mpl^2
    \label{matdom}
    \,.
\end{eqnarray}

Next, we use the well-known expressions for the mass and kinetic terms
of the supergravity Lagrangian, to estimate the magnitude of those
interaction terms in the Lagrangian which are proportional to
$\phi^2$, and are also proportional to either $\call\sub{kin}$ or
$\call\sub{mass}$ so that they can have a significant expectation
value.

Consider first terms proportional to $\call\sub{mass}$.  If the mass
term is that of a scalar field, it comes from the potential
\eq{vsugra} so that the factor $e^{K/\mpl^2}$ generically gives an
interaction term
\begin{eqnarray}
    \call\sub{massint} \sim \pm  \frac{\phi^2}{\mpl^2} \call\sub{mass}
    \,.
    \label{massint}
\end{eqnarray}
Considering instead chiral fermion, gaugino or the gravitino, all of
their mass terms come with a prefactor $e^{K/2\mpl^2}$, giving the
same interaction except for a factor $\frac12$.  In the case of scalar
particles, we noted before that higher-dimensional terms in $K$ could
lead to contributions comparable with the one coming from the
exponential factor, and an examination of the mass term for chiral
fermions leads to the same conclusion. With such terms, \eq{massint}
still provides an estimate of the interaction terms proportional to
$\phi^2$.  For massive gauge boson which acquires mass from the Higgs
mechanism, interaction term like (\ref{massint}) may also exist.  Mass
term of the gauge boson originates from the kinetic term of the Higgs
boson.  Thus, if the chiral multiplet of the Higgs $\Phi_H$ has a
coupling to $\Phi$ in the Kahler potential with the form $\sim
|\Phi_H|^2|\Phi|^2/\mpl^2$, interaction term like (\ref{massint})
shows up from the Kahler metric after the Higgs boson acquires the
VEV.

Now consider terms proportional to $\call\sub{kin}$.  If the kinetic
term is that of a scalar or of a chiral fermion, it arises because the
prefactor of the kinetic term is $K_{i\bar j}$.  This means that
Planck-suppressed interaction terms of the form $K=|\Phi_{\rm
gas}|^2|\Phi|^2/\mpl^2$ (with $\Phi_{\rm gas}$ being chiral multiplet
of the particles in the gas) will generate an interaction term
associated with the kinetic term
\begin{eqnarray}
    \call\sub{kinint} \sim \pm \frac{\phi^2}{\mpl^2} \call\sub{kin}
    \label{kinint}
    \,.
\end{eqnarray}
If instead the kinetic term is that of a gaugino or a gauge boson, a
similar conclusion follows if the gauge kinetic function contains a
term like $\sim \Phi^2/\mpl^2$.  Notice, however, that it is
model-dependent if such a term exists since it may vanish because of
some symmetry.

To summarize, we have found that the terms in the Lagrangian relevant
for determining the effective mass-squared of a field $\phi$ are
\begin{eqnarray}
    {\cal L} \supset \frac{1}{2}
    \partial_\mu \phi \partial^\mu \phi - \frac{1}{2} m_\phi^2 \phi^2
    +\( 1 - \lambda\sub{kin}\frac{\phi^2}{2\mpl^2}  \) \call\sub{kin} (\chi)
    +\( 1 - \lambda\sub{mass}\frac{\phi^2}{2\mpl^2}  \) \call\sub{mass} (\chi)
    \,,
    \label{L_tot}
\end{eqnarray}
with $\lambda\sub{kin}$ and $\lambda\sub{mass}$ expected to be of
order $\pm 1$.  Here $\chi$ stands for all of the fields corresponding
to particle species in the thermal bath, and $\phi$ is the scalar
field whose effective mass is to be evaluated.

With this preparation, the results are immediate.  Using
\eqss{kineqmass}{matdom}{L_tot} one finds during matter domination the
advertised result $m\sub{eff}^2 \sim \pm H^2$.  Using instead
\eqss{kineqmass}{raddom}{L_tot}, one finds during radiation domination
$|\meffs| \ll H^2$.  In fact, more precise study of $m\sub{eff}^2$ is
possible by calculating the expectation values of the relevant
operators in radiation- or matter-dominated universe.  For details,
see Appendix \ref{app:precise}.

As in the case of inflation, an additional contribution to the
effective mass-squared could come from terms in the superpotential.
Any such term would contribute to both the scalar and fermion fields
of a chiral supermultiplet. Consider in particular the term in
\eq{wbit}. The discussion in that case follows the lines of the one we
already gave for the case of inflation, except that we now need to
consider also the fermionic partner of $\Psi$ which we denote by
$\chi$. The terms contributing to the effective mass-squared of $\Phi$
is
\begin{eqnarray}
    \call \supset -
    \lambda\phi^2 
    \(  \frac{|\Psi|^4}{\mpl^2} + \frac{\bar\chi\chi}{\mpl} \)
    \label{callbit}
    \,,
\end{eqnarray}
with $\lambda\sim \pm 1$.  If the gas consists of $\Psi$ particles,
the first term is the relevant one, and will contribute
\begin{eqnarray}
    |\meffs| \supset   \mpl^{-2} \langle |\Psi|^4 \rangle
    \,,
\end{eqnarray}
which will be bigger than $H^2$ if $\langle |\Psi|^4 \rangle \gsim
\rho$.  If instead the gas consists of $\chi$ particles, \eq{callbit}
will contribute
\begin{eqnarray}
    |\meffs| \supset   \mpl^{-1} \langle \bar\chi \chi \rangle
    = (\mpl/m_\chi) \langle \call\sub{mass} \rangle /\mpl^2
    \,,
\end{eqnarray}
which is definitely much bigger than $H^2$.  However, as in the
discussion after \eqs{psiphi}{vbit}, these contributions to the
effective mass-squared of $\phi$ may be absent if $\phi$ is large,
because $\Psi$ and $\chi$ will then acquire large masses.  If large
enough, such masses will prevent the creation of $\Psi$ and $\chi$
particles, so that they will not be a component of the gas and will
not contribute to $\meffs$. Alternatively, the contribution may be
suppressed or forbidden by a symmetry.

To summarize, we have considered generic Planck-suppressed
interactions coming from the Kahler potential, and have found that
these generate a mass-squared of order $\pm H^2$ during
matter-domination but a much smaller one during radiation domination.
Just as in the case of inflation, further contributions to the
effective mass-squared may come from Planck-suppressed terms in $W$,
involving large slowly-varying fields (a generalization of the Higgs
effect).  In addition, a similar effect might come from rapidly
oscillating scalar fields which represent constituents of the gas.
Consider, for example, the following superpotential
\begin{eqnarray}
    W = y \Psi_1 \Psi_2 \Psi_3,
\end{eqnarray}
where $\Psi_i$ are the chiral superfields for the (scalar) particles
in the gas and $y$ is the coupling constant.  Then, taking account of
the effects of the Kahler potential, the scalar potential may contain
the term of the form
\begin{eqnarray}
    V = y^2 \left( 1 + \lambda \frac{|\Phi|^2}{\mpl^2} \right)
    |\Psi_1\Psi_2|^2 + \cdots.
\end{eqnarray}
From this potential, effective mass-squared of $\Phi$ receives a
contribution
\begin{eqnarray}
    m_{\rm eff}^2 = \frac{\lambda y^2 }{\mpl^2}
    \langle |\Psi_1|^2 \rangle \langle |\Psi_2|^2 \rangle  + \cdots.
\end{eqnarray}
If $y$ is small enough, $\Psi_i$ (approximately) obeys the equation of
motion of the massless particles.  Then, in the thermal bath, we
obtain $\langle |\Psi_i|^2 \rangle\sim T^2$ (see Appendix
\ref{app:precise}.)  If $y\ll 1$ (or $\lambda\ll 1$), $m_{\rm eff}$ is
smaller than $H$.  If $y\sim 1$, on the contrary, effective mass may
receive a contribution comparable to $H$. For the case of $y\sim 1$,
however, one should note that we would no longer be dealing with a gas
of free particles (ideal gas) since the interaction becomes very
strong.  Thus, in this case, the situation is different from the usual
'radiation- dominated' or `matter-dominated' epoch.

Finally, we consider the possibility that the mass of order $H$ during
matter domination might be suppressed. As with inflation, this is
certainly possible if the relevant field $\phi$ is a PNGB, and
suppression may also occur accidentally.  In contrast with inflation
though, it is difficult to find other ways of suppressing the mass,
because the scalar fields cannot be taken to be time-independent. As a
result, special choices of $K$ and $W$ are unlikely to work, while a
running mass will pass through zero only at a special epoch.

\section{Conclusion}

The considerations of this paper apply if there are scalar fields
whose true mass is less than the inflationary Hubble parameter
$H_*$. One expects that this will be the case if $H_*$ is of order the
gravitino mass (low-scale inflation) but not if it is much bigger.
Under that assumption, we considered the effective mass of a scalar
field $\phi$, whose true mass $m_\phi$ is indeed less than $H_*$, and
which has only Planck-suppressed interactions with the fields
responsible for the energy density of the Universe.

We have considered the era of inflation, and also the subsequent era
on the reasonable assumption that the Universe is then either matter-
or radiation-dominated.  For inflation and matter domination, we have
confirmed the received wisdom that the effective mass-squared is
generically of order $\pm H^2$.  We have also noted that this result
is quite difficult to avoid unless the relevant field is a PNGB,
especially in the case of matter domination. We then considered the
case of radiation domination, finding the perhaps unexpected result
that the effective mass-squared will be much less than $H^2$ if the
interaction among the particles consisting of the radiation are weak
enough.

These results have an important implication for the curvaton paradigm,
according to which the primordial density perturbation only at some
time after inflation ends, through the action of some `curvaton'
field. For this paradigm to be viable, the perturbation of the
curvaton field which is generated during inflation needs to be
maintained until the curvature perturbation is generated, which
probably requires \cite{dllr2} that the curvaton does {\em not} have a
mass of order $H$ between the end of inflation and the epoch when the
curvature perturbation is generated. This can be achieved by making
the curvaton a PNGB \cite{Dimopoulos:2003az}, but according to our
result it can also be achieved by having inflation give way promptly
to radiation domination.  Of course such early reheating, combined
with the high inflation scale that we are assuming, will lead to a
copious production of gravitinos, but as is well know these can be
sufficiently diluted by later entropy production coming from the decay
of a long-lived particle possible preceded by thermal inflation.

\paragraph{Acknowledgments}
T.M.\ would like to thank the organizers of COSMO03 workshop, where
this work was initiated, for their hospitality.  D.H.L.\ would like to
thank Tom Banks, Michael Dine, Renata Kallosh and Andrei Linde and
Mary K Gaillard for useful exchanges.  T.M. is supported by the
Grant-in-Aid for Scientific Research from the Ministry of Education,
Science, Sports, and Culture of Japan, No.\ 15540247.

\appendix

\section{A more precise analysis}
\label{app:precise}

In this appendix, we give a more precise analysis of the effective
mass-squared generated by a gas, making contact with the result of
finite-temperature effective field theory.  Here, we consider the
Lagrangian of the form \eq{L_tot}, where $\chi$ is the particle in the
thermal bath while $\phi$ is the scalar field whose effective mass is
to be evaluated.  Notice that, with the Planck-suppressed interactions
in the Kahler potential discussed in the previous section,
$|\lambda\sub{mass}|\sim 1$ and, if $\chi$ is in a chiral multiplet,
$|\lambda\sub{kin}|\sim 1$.

Consider first the case of a scalar field $\chi$; then,
$\call\sub{kin}$ and $\call\sub{mass}$ are given by
\begin{eqnarray}
    \call\sub{kin} (\chi)  &=&
    \frac12 \partial_\mu \chi \partial^\mu \chi \,,
    \label{L_kin}
    \\
    \call\sub{mass} (\chi) &=& -\frac12 m_\chi^2 \chi^2 \,.
    \label{L_mass}
\end{eqnarray}
We expand the field operator $\chi$ using the creation and
annihilation operators $a_{\bf p}$ and $a_{\bf p}^\dagger$.  Here, we
use the box normalization of the wave functions using the box with the
volume $V=L^3$.  Then, we obtain
\begin{eqnarray}
    \chi (x) = \sum_{\bf p}
    \frac{1}{\sqrt{2E_{\bf p} V}}
    \left( a_{\bf p} e^{-ipx} + a_{\bf p}^\dagger e^{ipx} \right)
    \label{chi}
    \,,
\end{eqnarray}
where the four momentum $p=(E_{\bf p},{\bf p})$ obeys the on-shell
condition $k^2=m_\chi^2$.  (Here, we expect that the time scale to
realize the equilibrium is much shorter than the cosmic time scale
$\sim H^{-1}$, and hence we neglect the effect of red-shift in
evaluating the expectation values.)  In addition, spatial components
of the momentum is given in the form ${\bf p}= \frac{2\pi}{L}(n_x,
n_y, n_z)$ with $n_x$, $n_y$, and $n_z$ being integers.  Notice that,
with this normalization, $[a_{\bf p}, a_{\bf p'}^\dagger]=\delta_{\bf
pp'}$.

Substituting Eq.\ (\ref{chi}) into Eqs.\ (\ref{L_kin}) and
(\ref{L_mass}), expectation values of ${\cal L}_{\rm kin}$ and ${\cal
L}_{\rm mass}$ are given by
\begin{eqnarray}
    \langle {\cal L}_{\rm kin} \rangle =
    \frac{1}{V} \sum_{\bf p} \frac{k^2}{2E_{\bf p}}
    \langle a_{\bf p}^\dagger a_{\bf p} \rangle
    \,, ~~~
    \langle {\cal L}_{\rm mass} \rangle =
    - \frac{1}{V} \sum_{\bf p} \frac{m_\chi^2}{2E_{\bf p}}
    \langle a_{\bf p}^\dagger a_{\bf p} \rangle
    \,.
\end{eqnarray}
With the replacement $\frac{1}{V}\sum_{\bf
p}\rightarrow\int\frac{d^3{\bf p}}{(2\pi)^3}$ and $k^2=m_\chi^2$, the
above formulae become
\begin{eqnarray}
    \langle {\cal L}_{\rm kin} \rangle
    = - \langle {\cal L}_{\rm mass} \rangle
    = \frac{1}{2} m_\chi^2
    \int \frac{d^3{\bf p}}{(2\pi)^3}
    \frac{\langle n_\chi ({\bf p})\rangle}{E_{\bf p}}
    \,,
\end{eqnarray}
where $\langle n_\chi ({\bf p})\rangle=\langle a_{\bf p}^\dagger
a_{\bf p}\rangle$ is the expectation value of the number density of
$\chi$ with the momentum ${\bf p}$.  Using the relation $\langle {\cal
L}_{\rm kin}\rangle= - \langle {\cal L}_{\rm mass} \rangle$, we obtain
\begin{eqnarray}
    m\sub{eff}^2 =  \frac{1}{M_{\rm P}^2}
    \( \lambda\sub{kin} - \lambda\sub{mass} \)
    \langle \call\sub{kin} \rangle
    \,.
    \label{dm^2}
\end{eqnarray}
Also, the energy density is
\begin{eqnarray}
    \rho
    =
    \int \frac{d^3{\bf p}}{(2\pi)^3}
    E_{\bf p} \langle n_\chi ({\bf p})\rangle
    \,.
\end{eqnarray}

During matter domination, the random motion of the gas of $\chi$
particles is non-relativistic, which means that $\langle n_\chi ({\bf
p})\rangle$ is suppressed unless $|{\bf p}|\ll m_\chi$, and
\begin{eqnarray}
    \langle \call\sub{kin} \rangle \simeq \frac{1}{2} \rho_\chi
    \simeq \frac{1}{2} m_\chi
    \int \frac{d^3{\bf p}}{(2\pi)^3}
    \langle n_\chi ({\bf p})\rangle
    \,.
\end{eqnarray}
Denoting the energy fraction of $\chi$ as $\Omega_\chi$, $\Delta
m_\phi^2$ is given by
\begin{eqnarray}
    m\sub{eff}^2
    = \( \lambda\sub{kin} - \lambda\sub{mass} \)
    \frac{\rho_\chi}{2\mpl^2}
    = \frac{3}{2} \( \lambda\sub{kin} - \lambda\sub{mass} \)
    \Omega_\chi H^2
    \label{md}
    \,.
\end{eqnarray}
Thus, during a matter-dominated era when the non-relativistic gas of
$\chi$ dominates the Universe, an effective mass of $O(\pm H^2)$ is
generated, provided that $\lambda\sub{kin}$ and $\lambda\sub{mass}$
are of $O(1)$.

During radiation domination, the random motion of the gas of $\chi$
particles is non-relativistic, which means that $\langle n_\chi ({\bf
p})\rangle$ is suppressed unless $E_{\bf p}\simeq p \ll m_\chi$. Then
$\rho\ll \langle \call\sub{mass} \rangle$ and $m\sub{eff}^2 \ll H^2$.

We note here the special case that the radiation is thermalized with
negligible chemical potential. Then
\begin{eqnarray}
    \langle n_\chi ({\bf p})\rangle =
    \frac{1}{e^{|{\bf p}|/T}-1}
    \,,
\end{eqnarray}
and
\begin{eqnarray}
    \langle {\cal L}_{\rm kin} \rangle
    = \frac{1}{24} m_\chi^2 T^2
    \,,
\end{eqnarray}
and using $\rho_\chi=\frac{\pi^2}{30}T^4$,
\begin{eqnarray}
    m\sub{eff}^2 = \frac{15}{4\pi^2}
    \( \lambda\sub{kin} - \lambda\sub{mass} \)
    \Omega_\chi
    \frac{m_\chi^2}{T^2} H^2
    \,.
    \label{rd}
\end{eqnarray}
(Notice that the term proportional to $\lambda\sub{mass}$ is
consistent with the thermal mass-squared given in \cite{Dolan:qd}.)
In this case, $m\sub{eff}^2/H^2$ is of order $\sim m_\chi^2/T^2$ or
smaller.

With $\chi$ being a fermion, there is no qualitative change in the
above results.  Taking $\chi$ to be a spin $\frac{1}{2}$ fermion, we
have
\begin{eqnarray}
    \call\sub{kin} (\chi) &=& i\bar\chi \dslash \chi
    \,,
    \\
    \call \sub{mass} (\chi) &=& - m_\chi \bar\chi \chi
    \,.
    \label{L_tot2}
\end{eqnarray}
Then, the $\chi$ field is expanded as
\begin{eqnarray}
    \chi (x) = \sum_{{\bf p}, s}
    \frac{1}{\sqrt{2E_{\bf p} V}}
    \left( b_{{\bf p}, s} u_{{\bf p}, s} e^{-ipx}
        + d_{{\bf p}, s}^\dagger v_{{\bf p}, s} e^{ipx} \right)
    \label{chi(dirac)}
    \,,
\end{eqnarray}
where $s$ is the spin index.  Here, $u_{{\bf p}, s}$ and $v_{{\bf p},
s}$ are Dirac spinors, and $b_{{\bf p}, s}$ and $d_{{\bf p},
s}^\dagger$ are creation and annihilation operators obeying $\{b_{{\bf
p}, s}, b_{{\bf p'}, s'}^\dagger\}=\{d_{{\bf p}, s}, d_{{\bf p'},
s'}^\dagger\}=\delta_{{\bf p}{\bf p'}}\delta_{s, s'}$.  (The following
argument does not change even when $\chi$ is a Majorana particle.)
Dirac equation gives $\langle\call\sub{kin}\rangle
=-\langle\call\sub{mass}\rangle$ and hence, with the Lagrangian given
in Eq.\ (\ref{L_tot}), we also obtain the same expression of $\Delta
m_\phi^2$ as given in Eq.\ (\ref{dm^2}).  For a Dirac fermion,
expectation value of ${\cal L}_{\rm kin}$ is given by
\begin{eqnarray}
    \langle {\cal L}_{\rm kin} \rangle =
    m_\chi^2 \int \frac{d^3 {\bf p}}{(2\pi)^3}
    \frac{\langle n_{\chi} ({\bf p}) \rangle
    + \langle n_{\bar{\chi}} ({\bf p}) \rangle}
    {E_{\bf p}}
    \,,
\end{eqnarray}
where $n_{\chi} ({\bf p})=\sum_s b_{{\bf p}, s}^\dagger b_{{\bf p},
s}$ and $n_{\bar{\chi}} ({\bf p})=\sum_s d_{{\bf p}, s}^\dagger
d_{{\bf p}, s}$, while the energy density is
\begin{eqnarray}
    \rho_\chi =
    \int \frac{d^3 {\bf p}}{(2\pi)^3} E_{\bf p}
    \left[ \langle n_{\chi} ({\bf p}) \rangle
        + \langle n_{\bar{\chi}} ({\bf p}) \rangle \right]
    \,.
\end{eqnarray}
In the non-relativistic case $E_{\bf p}$ in the above expressions can
be replaced by $m_\chi$ again giving \eq{md} except for the factor
$\frac{1}{2}$.  In addition, in the relativistic case, $\langle {\cal
L}_{\rm kin} \rangle =\frac{1}{6}m_\chi^2T^2$ taking account of the
spin degrees of freedom, and hence $m\sub{eff}^2$ becomes
\begin{eqnarray}
    m\sub{eff}^2 = \frac{30}{7\pi^2}
    \( \lambda\sub{kin} - \lambda\sub{mass} \)
    \Omega_\chi
    \frac{m_\chi^2}{T^2} H^2
    \,.
\end{eqnarray}
Thus, the effective mass induced by relativistic fermions is again
much smaller than $H$.


\begin{thebibliography}{99}
    
\bibitem{Dolan:qd}
    L.~Dolan and R.~Jackiw,
    Phys.\ Rev.\ D {\bf 9} (1974) 3320.
    
\bibitem{myhigher}
    D.~H.~Lyth,
    Phys.\ Lett.\ B {\bf 419} (1998) 57.
    
\bibitem{hwmass}
    H.~P.~Nilles, M.~Olechowski and M.~Yamaguchi,
    Phys.\ Lett.\ B {\bf 415} (1997) 24.
    
\bibitem{inflation}
    B.~A.~Ovrut and P.~J.~Steinhardt,
    Phys.\ Lett.\ B {\bf 133} (1983) 161;
    G.~D.~Coughlan, R.~Holman, P.~Ramond and G.~G.~Ross,
    Phys.\ Lett.\ B {\bf 140} (1984) 44;
    M.~Dine, W.~Fischler and D.~Nemeschansky,
    Phys.\ Lett.\ B {\bf 136} (1984) 169.
    
\bibitem{cllsw}
    E.~J.~Copeland, A.~R.~Liddle, D.~H.~Lyth, E.~D.~Stewart and D.~Wands,
    Phys.\ Rev.\ D {\bf 49} (1994) 6410.
    
\bibitem{drt}
    M.~Dine, L.~Randall and S.~Thomas,
    Phys.\ Rev.\ Lett.\  {\bf 75} (1995) 398;
    Nucl.\ Phys.\ B {\bf 458} (1996) 291.
    
\bibitem{pqflaton}
    E.~J.~Chun, D.~Comelli and D.~H.~Lyth,
    Phys.\ Rev.\ D {\bf 62} (2000) 095013;
    E.~J.~Chun, H.~B.~Kim and D.~H.~Lyth,
    Phys.\ Rev.\ D {\bf 62} (2000) 125001.
    
\bibitem{Affleck:1984fy}
    I.~Affleck and M.~Dine,
    Nucl.\ Phys.\ B {\bf 249} (1985) 361.
    
\bibitem{kari}
    A.~Mazumdar and A.~Perez-Lorenzana,
    arXiv:hep-ph/0311106;
    K.~Enqvist, S.~Kasuya and A.~Mazumdar,
    arXiv:hep-ph/0311224.

\bibitem{treview}
    D.~H.~Lyth and A.~Riotto,
    Phys.\ Rept.\  {\bf 314} (1999) 1.
    
\bibitem{book}
    A.~R.~Liddle and D.~H.~Lyth,
    {\sl Cosmological inflation and large-scale structure,},
    Cambridge, UK: Univ. Pr. (2000).
    
\bibitem{Lyth:2001nq}
    D.~H.~Lyth and D.~Wands,
    Phys.\ Lett.\ B {\bf 524} (2002) 5.
    
\bibitem{Moroi:2001ct}
    T.~Moroi and T.~Takahashi,
    Phys.\ Lett.\ B {\bf 522} (2001) 215
    [Erratum-ibid.\ B {\bf 539} (2002) 303].
    
\bibitem{Enqvist:2001zp}
    K.~Enqvist and M.~S.~Sloth,
    Nucl.\ Phys.\ B {\bf 626} (2002) 395.
    
\bibitem{wmapspergel}
    D.~N.~Spergel {\it et al.},
    astro-ph/0302209.
    
\bibitem{wmapsdss}
    M.~Tegmark {\it et al.}  [SDSS Collaboration],
    astro-ph/0310723.
    
\bibitem{dllr2}
    K.~Dimopoulos, G.~Lazarides, D.~Lyth and R.~Ruiz de Austri,
    JHEP {\bf 0305} (2003) 057.
    
\bibitem{gauge}
    N.~Arkani-Hamed, H.~C.~Cheng, P.~Creminelli and L.~Randall,
    Phys.\ Rev.\ Lett.\  {\bf 90} (2003) 221302;
    JCAP {\bf 0307} (2003) 003.
    
\bibitem{banks03}
    T.~Banks, M.~Dine, P.~J.~Fox and E.~Gorbatov,
    JCAP {\bf 0306} (2003) 001.
    
\bibitem{joebook}
    J.~Polchinski,
    {\sl ``String Theory. Vol. 2: Superstring Theory And Beyond,''},
    Cambridge, UK: Univ. Pr. (1998).
    
\bibitem{fofphi}
    T.~Banks and M.~Dine,
    Phys.\ Rev.\ D {\bf 50} (1994) 7454;
    T.~Banks and M.~Dine,
    Nucl.\ Phys.\ B {\bf 479} (1996) 173.
    
\bibitem{thooft}
    G.~'t Hooft,
    in ``Recent Developments in Gauge Theories''
    (1980, Plenum Publishing Co.) 135.
    
\bibitem{p03stringph}
    K.~Dimopoulos and D.~H.~Lyth,
    arXiv:hep-ph/0209180;
    D.~H.~Lyth,
    arXiv:hep-th/0311040.
    
\bibitem{ewansugra}
    E.D. Stewart,
    Phys.\ Rev.\ D {\bf 51} (1995) 6847.
    
\bibitem{wein3}
    S.~Weinberg,
    {\sl The Quantum Theory Of Fields.  Vol. 3: Supersymmetry},
    Cambridge, UK: Univ. Pr. (2000).
    
\bibitem{Gaillard:1993es}
    M.~K.~Gaillard and V.~Jain,
    Phys.\ Rev.\ D {\bf 49} (1994) 1951.
    
\bibitem{Bagger:1995ay}
    J.~Bagger, E.~Poppitz and L.~Randall,
    Nucl.\ Phys.\ B {\bf 455} (1995) 59.
    
\bibitem{Stewart:2000pa}
    J.~D.~Cohn and E.~D.~Stewart,
    Phys.\ Lett.\ B {\bf 475} (2000) 231;
    E.~D.~Stewart and J.~D.~Cohn,
    Phys.\ Rev.\ D {\bf 63} (2001) 083519.
    
\bibitem{cdl}
    E.\ J.\ Chun, K.\ Dimopoulos and D.\ H.\ Lyth,
    arXiv:hep-ph/0402059.
    
\bibitem{naturalinf}
    K.~Freese, J.~A.~Frieman and A.~V.~Olinto,
    Phys.\ Rev.\ Lett.\  {\bf 65} (1990) 3233;
    F.~C.~Adams, J.~R.~Bond, K.~Freese, J.~A.~Frieman and A.~V.~Olinto,
    Phys.\ Rev.\ D {\bf 47} (1993) 426.
    
\bibitem{Cohn:2000hc}
    J.~D.~Cohn and E.~D.~Stewart,
    Phys.\ Lett.\ B {\bf 475} (2000) 231.
    
\bibitem{kaplan}
    D.~E.~Kaplan and N.~J.~Weiner,
    arXiv:hep-ph/0302014.
    
\bibitem{renata}
    J.~P.~Hsu, R.~Kallosh and S.~Prokushkin,
    JCAP {\bf 0312} (2003) 009.
    
\bibitem{ars}
    J.~A.~Adams, G.~G.~Ross and S.~Sarkar,
    Phys.\ Lett.\ B {\bf 391}, 271 (1997).
    
\bibitem{Dimopoulos:2003az}
    K.~Dimopoulos, D.~H.~Lyth, A.~Notari and A.~Riotto,
    JHEP {\bf 0307} (2003) 053.
    
\bibitem{string-inspired}
    F.~Koyama, Y.~Tachikawa and T.~Watari,
    arXiv:hep-th/0311191;
    J.~P.~Hsu and R.~Kallosh,
    arXiv:hep-th/0402047.
    
\bibitem{shift-chaotic}
    M.~Kawasaki, M.~Yamaguchi and T.~Yanagida,
    Phys.\ Rev.\ Lett.\  {\bf 85} (2000) 3572;
    M.~Yamaguchi and J.~Yokoyama,
    Phys.\ Rev.\ D {\bf 63} (2001) 043506;
    M.~Yamaguchi,
    Phys.\ Rev.\ D {\bf 64} (2001) 063502;
    M.~Yamaguchi and J.~Yokoyama,
    Phys.\ Rev.\ D {\bf 68} (2003) 123520.
    
\bibitem{strax}
    T.~Banks and M.~Dine,
    Nucl.\ Phys.\ B {\bf 505} (1997) 445;
    T.~Banks, M.~Dine and M.~Graesser,
    Phys.\ Rev.\ D {\bf 68} (2003) 075011.
    
\bibitem{glm}
    M.~K.~Gaillard, D.~H.~Lyth and H.~Murayama,
    Phys.\ Rev.\ D {\bf 58}, 123505 (1998).
    
\bibitem{kklmmt}
    S.~Kachru, R.~Kallosh,
    A.~Linde, J.~Maldacena, L.~McAllister and S.~P.~Trivedi,
    JCAP {\bf 0310} (2003) 013.
    
\bibitem{henry}
    H.~Firouzjahi and S.~H.~H.~Tye,
    arXiv:hep-th/0312020.
    
\bibitem{adhoc}
    A.\ D.\ Linde, {\sl Particle Physics and Inflationary Cosmology},
    Harwood Academic, Switzerland (1990);
    K.~A.~Olive,
    Phys.\ Rept.\  {\bf 190} (1990) 307;
    H.~Murayama, H.~Suzuki, T.~Yanagida and J.~Yokoyama,
    Phys.\ Rev.\ D {\bf 50} (1994) 2356.
    
\bibitem{gmo}
    M.~K.~Gaillard, H.~Murayama and K.~A.~Olive,
    Phys.\ Lett.\ B {\bf 355} (1995) 71.
    
\bibitem{running1}
    E.~D.~Stewart,
    Phys.\ Lett.\ B {\bf 391}, 34 (1997).
    
\bibitem{running2}
    E.~D.~Stewart,
    Phys.\ Rev.\ D {\bf 56}, 2019 (1997).
    
\bibitem{clm}
    L.~Covi, D.~H.~Lyth and A.~Melchiorri,
    Phys.\ Rev.\ D {\bf 67} (2003) 043507.
    
\bibitem{ks}
    K.~Kadota and E.~D.~Stewart,
    JHEP {\bf 0307}, 013 (2003).
    
\bibitem{Halyo:1996pp}
    E.~Halyo,
    Phys.\ Lett.\ B {\bf 387} (1996) 43.
    
\bibitem{Binetruy:1996xj}
    P.~Binetruy and G.~R.~Dvali,
    Phys.\ Lett.\ B {\bf 388} (1996) 241.
    
\bibitem{km}
    C.~F.~Kolda and J.~March-Russell,
    Phys.\ Rev.\ D {\bf 60} (1999) 023504.
    
\end{thebibliography}
\end{document}